\documentclass[aps,prb,twocolumn,amsmath,amssymb,floatfix]{revtex4-1}
\usepackage{amsmath}
\usepackage{amssymb}
\usepackage{amsfonts}
\usepackage{listings}
\usepackage{enumerate}
\usepackage{latexsym}
\usepackage{multirow}
\usepackage{graphicx}
\usepackage{braket}
\usepackage[pdftex,colorlinks=false]{hyperref}
\usepackage{xcolor, colortbl}
\usepackage{verbatim}
\usepackage{times}
\usepackage{array, makecell}
\usepackage[export]{adjustbox}
\usepackage{tabularx}
\usepackage{mathtools}

\newcolumntype{M}[1]{>{\centering\arraybackslash}m{#1}} %for centering in tables.

\newcolumntype{N}{@{}m{0pt}@{}}

\newcommand{\beq}{\begin{equation}}
\newcommand{\eneq}{\end{equation}}
\newcommand{\bs}[1]{\boldsymbol{#1}}

\def\be{\begin{equation}}
\def\ee{\end{equation}}
\def\ba{\begin{eqnarray}}
\def\ea{\end{eqnarray}}

\def\R{{\rm Re}}
\def\Z{\mathbb{Z}}
\def\C{\mathbb{C}}

\renewcommand{\vec}{\bs}

\def\beq{\begin{equation}}
\def\eeq{\end{equation}}
\def\barray{\begin{eqnarray}}
\def\earray{\end{eqnarray}}

\font\upright=cmu10 scaled\magstep1
\def\stroke{\vrule height8pt width0.4pt depth-0.1pt}

\def\Zmath{\mathbb{Z}}
\def\Qmath{\vcenter{\hbox{\upright\rlap{\rlap{Q}\kern
                   3.8pt\stroke}\phantom{Q}}}}
\def\Nmath{\vcenter{\hbox{\upright\rlap{I}\kern 1.7pt N}}}
\def\Cmath{\vcenter{\hbox{\upright\rlap{\rlap{C}\kern
                   3.8pt\stroke}\phantom{C}}}}
\def\Rmath{\vcenter{\hbox{\upright\rlap{I}\kern 1.7pt R}}}
\def\Z{\ifmmode\Zmath\else$\Zmath$\fi}
\def\Q{\ifmmode\Qmath\else$\Qmath$\fi}
\def\N{\ifmmode\Nmath\else$\Nmath$\fi}
\def\C{\ifmmode\Cmath\else$\Cmath$\fi}
\def\R{\ifmmode\Rmath\else$\Rmath$\fi}

\input{epsf}

\newcounter{defcounter}
\begin{document}

%\tolerance 10000
%
%
%
%\widowpenalty10000
%\clubpenalty10000

\title{Flat bands with fragile topology through superlattice engineering on single-layer graphene}

\author{
Anastasiia~Skurativska}
\affiliation{
 Department of Physics, University of Zurich, Winterthurerstrasse 190, 8057 Zurich, Switzerland
}
\author{Stepan~S.~Tsirkin
}
\affiliation{
 Department of Physics, University of Zurich, Winterthurerstrasse 190, 8057 Zurich, Switzerland
}
\author{Fabian~D~Natterer
}
\affiliation{
 Department of Physics, University of Zurich, Winterthurerstrasse 190, 8057 Zurich, Switzerland
}
\author{
Titus~Neupert}
\affiliation{
 Department of Physics, University of Zurich, Winterthurerstrasse 190, 8057 Zurich, Switzerland
}
\author{
Mark~H~Fischer}
\affiliation{
 Department of Physics, University of Zurich, Winterthurerstrasse 190, 8057 Zurich, Switzerland
}

\begin{abstract}
    `Magic'-angle twisted bilayer graphene has received a lot of interest due to its flat bands with potentially non-trivial topology that lead to intricate correlated phases.
    A spectrum with flat bands, however, does not require a twist between multiple sheets of van der Waals materials, but rather can be realized with the application of an appropriate periodic potential.
    Here, we propose the imposition of a tailored periodic potential onto a single graphene layer through local perturbations that could be created via lithography or adatom manipulation, which also results in an energy spectrum featuring flat bands. 
    Our first-principle calculations for an appropriate decoration of graphene with adatoms indeed show the presence of flat bands in the spectrum. 
    Furthermore,  we reveal the topological nature of the flat bands through a symmetry-indicator analysis. 
    This non-trivial topology manifests itself in corner-localized states with a filling anomaly as we show using a tight-binding model.
    Our proposal of a single decorated graphene sheet provides a new versatile route to study correlated phases in topologically non-trivial, flat band structures.

\end{abstract}

\date{\today}

\maketitle

%\section*{Introduction}

Engineering the desired functionality of a material through nanostructuring has proven a powerful approach that is particularly well suited for two-dimensional (2D) materials. Often, the goal of such engineering involves shifting spectral weight to the Fermi level or increasing the coupling of the electrons to other degrees of freedom, such as light for improved optoelectronic properties~\cite{yan:2020} or phonons for superconductivity~\cite{allan:2017}. Nanostructuring can be achieved in a variety of ways: standard cleanroom techniques, such as electron beam lithography~\cite{grigorescu:2009}, photolithography, or focused ion beam lithography~\cite{genet:2007}, allow to realize patterns of a few nanometers in size. 

Moir\'e engineering, in other words using the potential landscape from a Moir\'e lattice that emerges due to a finite twist angle between two 2D lattices, has attracted a lot of attention in recent years.
This attention stems in large part from the discovery of extremely flat bands for certain, very small twist angles, referred to as `magic' angles, in twisted bilayer graphene (TBG)~\cite{mcdonald:2011, cao:2018a, herrero:2018}. Ideally, these bands concentrate spectral weight around the Fermi level, are separated by a gap from other bands in the spectrum, and potentially possess non-trivial topology with intriguing implications for the many-body ground states~\cite{zou:2020,bernevig:2020}. Moreover, as a result of the flatness of these bands the electron-electron coupling becomes the dominant interaction. Consequently, this system shows correlated insulator states at integer fillings~\cite{cao:2018a} and superconductivity in between~\cite{herrero:2018,efetov:2019}. Finally, even quantum anomalous Hall phases, driven by the strong interactions, have been observed~\cite{Wang:2021,pierce:2021,serlin:2020}.

\begin{figure*}[t]
  \centering
  \includegraphics[width=0.95\linewidth]{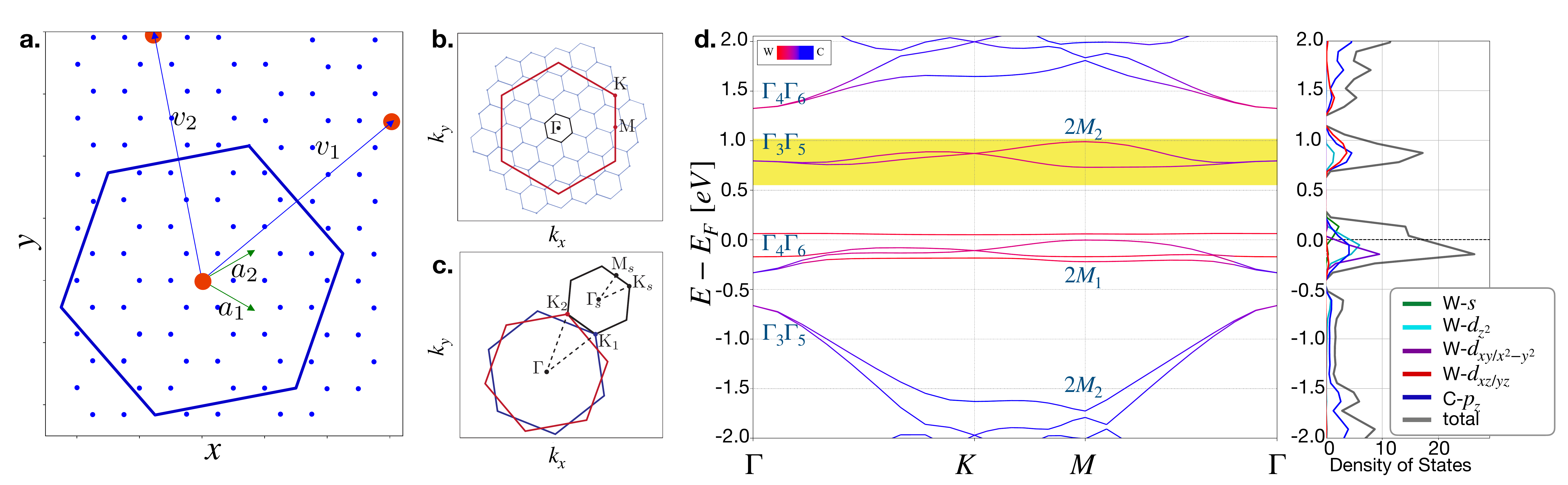}
  \caption{\label{fig:unit_cell} {\bf System setup and electronic structure.} ({\bf a}) The unit cell consisting of a graphene monolayer with adatoms placed in the center of the unit cell as indicated by the red circle. The lattice vectors ($\vec{v_1}$,$\vec{v_2}$) are related to the lattice vectors of pure graphene ($\vec{a_1}$, $\vec{a_2}$) by Eq.~(\ref{eq:relation_lat_vecs}) with $(n,m) = (-1, 2)$. ({\bf b}) The Brillouin zone (BZ) of monolayer graphene shown in red and the reduced BZ of our setup depicted in black. The $K$ and $K'$ points of graphene map to the $\Gamma$ point in the reduced BZ of the superstructure. ({\bf c}) For comparison, the BZ of TBG (in black) tilted with respect to the two BZs of graphene twisted by a small angle (in blue and red). ({\bf d}) The DFT spectrum of graphene with tungsten adatoms. Irreducible representations of the bands at high-symmetry points are indicated in the figure. Right side shows the density of states projected onto the adatom and carbon $p_z$ orbitals, respectively.}
\end{figure*}

More specifically, for a discrete set of angles, the Moir\'e pattern creates a commensurate hexagonal supercell and the electronic structure can be described again in terms of Bloch states. The superstructure significantly affects the tunneling of electrons between the two graphene layers and the hybridization of electronic states near the two graphene Dirac-cones results in a decrease of the Fermi velocity at the charge neutrality point and the formation of nearly flat bands in the spectrum.
The twist angle in bilayer graphene directly controls the size of the Moir\'e supercell and thus acts as the tuning parameter for the band flattening.

Motivated by magic-angle TBG, many more systems were proposed, where a small twist angle between stacked layers, such as other multilayer graphene heterostructures or transition metal dichalcogenides, is used to create novel correlated phases. Furthermore, Moir\'e engineering on a single Dirac cone of a topological-insulator surface state was recently discussed~\cite{cano:2020, wang:2020}.
Besides the restriction on the types of periodic potentials that can be realized with Moir\'e pattern, however, producing samples with predefined twist angle and sufficient homogeneity is a further intricate experimental challenge~\cite{uri:2020, benschop:2020tmp}. There is thus an ongoing search for other systems with electronic properties similar to the ones of magic-angle TBG~\cite{lee:2020}, but with more control over the design. Such systems provide novel platforms to study the physics of correlated electrons in topologically non-trivial bands.

In this work, we propose an alternative approach for creating topologically non-trivial flat bands in a single graphene sheet by the application of a periodic potential. In particular, using first-principle calculations, we investigate a single layer of graphene decorated with a periodic, $C_6$-symmetric distribution of adatoms, see Fig.~\ref{fig:unit_cell}a. While such an approach allows for high control over the applied potential through an adatom superlattice via atom manipulation using scanning tunnelling microscopy~\cite{brar:2011,wang:2013,wyrick:2016}, our principle is amenable also to artificial graphene~\cite{gomes:2012,Drost:2017} or engineered lattices~\cite{Drost:2017,Yan:2019,khajetoorians:2019}, and via nanofabrication to graphene~\cite{dyck:2017}.

Within our first-principles calculations, we indeed find flat bands separated by gaps from other bands in the spectrum of the system. Employing the recently introduced framework of topological quantum chemistry~\cite{TQC:2017}, we further reveal the fragile topological nature of these bands. Bands with this type of topology stand in between strong topological and trivial phases, based on the topological robustness against addition of trivial degrees of freedom. Specifically, the main characteristic of bands with fragile topology is that they can be trivialised by addition of bands permitting an atomic limit below the Fermi level \cite{TQC:2017,watanabe:2017,po:2018}.

Systems with fragile topology protected by $n$-fold rotation symmetry often feature a filling anomaly: In open boundary conditions, the system possesses $n$ degenerate in-gap states, which are only partially occupied at charge neutrality~\cite{huges:2019}. This implies a degeneracy of the many-body ground state in the thermodynamic limit protected by rotation symmetry. We illustrate this bulk-boundary correspondence employing a tight-binding calculation of a $C_6$-symmetric flake, where the filling anomaly manifests itself in an excess charge accumulation in the corners of the flake, suitable for experimental discovery.

\section*{Model system and flat bands}

Conceptually, the electronic structure of both TBG and our approach can be understood starting from that of a single layer of graphene.
The band structure is characterized by two Dirac cones around the $K$ and $K'$ points in the Brillouin zone (BZ). For TBG with discrete commensurate twist angles, the resulting periodic superstructure allows for a momentum-space description in a reduced BZ. As shown in Fig.~\ref{fig:unit_cell}c, this reduced BZ can be geometrically constructed directly in momentum space, by twisting the two original BZs resulting in a reduced BZ with new $K$ and $K'$ points stemming from the $K$ points from the two individual layers, $K_1$ and $K_2$. For small twist angles, the tunneling between the two graphene layers that hybridizes the bands around $K_1$ and $K_2$ becomes comparable to the band width of the respective bands in the reduced BZ, resulting in flat bands over the whole reduced BZs. Note that there are two time-reversal related BZs, one from the $K$ and one from the $K'$ points of the individual graphene layers. For the small angles required for flat bands, the resulting unit cell in real space contains thousands of carbon atoms.

Inspired by the construction in TBG, we build in the following flat bands starting again from a single graphene sheet, but enlarging the unit cell using nanostructuring. We consider a  superlattice potential arising from adatoms placed in the hollow sites (H) of the graphene lattice, meaning in the center of the C hexagon, with a periodicity described by the superlattice vectors $\vec{v}_1$ and $\vec{v}_2$, see Fig.~\ref{fig:unit_cell}a. In consequence and contrast to the TBG case, we therefore only have a single original $K$ and $K'$ point. We choose the superlattice vectors in such a way, that both $K$ and $K'$ are mapped to the $\Gamma$ point of the reduced BZ. Crucially, this allows for a strong hybridization of the original with the adatom bands. 
The lattice vectors leading to such a configuration are given by 
\begin{equation}\label{eq:relation_lat_vecs}
    \vec{v}_1 = n \vec{a}_1 + (3m+n) \vec{a}_2\, , \\
%    &\vec{v_2} = R_6 \vec{v_1}\,,
\end{equation}
where $n,m \in \mathbb{Z}$, $\vec{a}_1$ and $\vec{a}_2$ are the lattice vectors for graphene, and $\vec{v}_2$ is related to $\vec{v}_1$ by a 60-degrees rotation.
For concreteness, we use in the following $(n,m) = (-1, 2)$.
The BZ resulting from this construction is shown in Fig.~\ref{fig:unit_cell}b.

As a guiding principle, we choose non-magnetic transition-metal adatoms, which, as we will discuss below, generically lead to flat bands that are topologically non-trivial due to their $d$ orbitals. We further require the candidate adatoms to be sufficiently stable at the H site~\cite{nakada:2011}. As a figure of merit, we consider the ratio of spectral gap to the closest C-based bands and the band width. Finally focusing on situations, where the flat bands are separated from other bands related to the adatoms, these guiding principles result in a set of most promising adatoms: W, Ta, and Ru. Figure~\ref{fig:unit_cell}d shows the band structure that results from the above construction with tungsten adatoms obtained using first-principles calculations (see Methods for details). For these calculations, the graphene lattice is oriented in the $x$-$y$ plane and the tungsten atoms are relaxed to their equilibrium position in $z$ direction over the H site.
For completeness, Tab.~\ref{tab:table-elements} summarizes the band width and gaps to the nearest C-based bands of tungsten and all other considered adatoms (see Methods for their corresponding DFT spectra).

\begin{table}[b]
    \begin{tabular}{c|cccccccccc}
    \hline
    $\text{Element}$&$\text{W}$&$\text{Ir}$&$\text{Cr}$&$\text{Mo}$&$\text{Rh}$&$\text{Ta}$&$\text{Ru}$&$\text{Re}$&$\text{Os}$&$\text{Nb}$\\
    \hline
    $\text{Width [eV]}$&$0.26$&$0.56$&$0.21$&$0.30$&0.39&0.27&0.28&0.24&0.30&0.30\\
    %\hline
    $\text{Gap [eV]}$&$0.34$&$0.26$&$0.26$&$0.36$&0.30&0.31&0.38&0.39&0.37&0.33\\
    \hline
    \end{tabular}
    \caption{Summary of adatoms placed in H sites and that form flat bands, featuring fragile topology with band width and spectral gap to the nearest C bands indicated. }
    \label{tab:table-elements}
\end{table}

In the following we focus on the flat bands highlighted in Fig.~\ref{fig:unit_cell}d, assuming that the chemical potential through the graphene channel can be directly controlled by back gating~\cite{Novoselov:2004}.
The orbital projected density of states on the right side of Fig.~\ref{fig:unit_cell}d shows that these flat bands are originating from the hybridization of the C $p_z$ orbitals of graphene and the $d_{xz,yz}$ orbitals of tungsten. As we will discuss in the following, this is crucial for the topological properties and needs to be taken into account when constructing a minimal tight-binding model.

\begin{figure}[t]
  \centering
  \includegraphics[width=\linewidth]{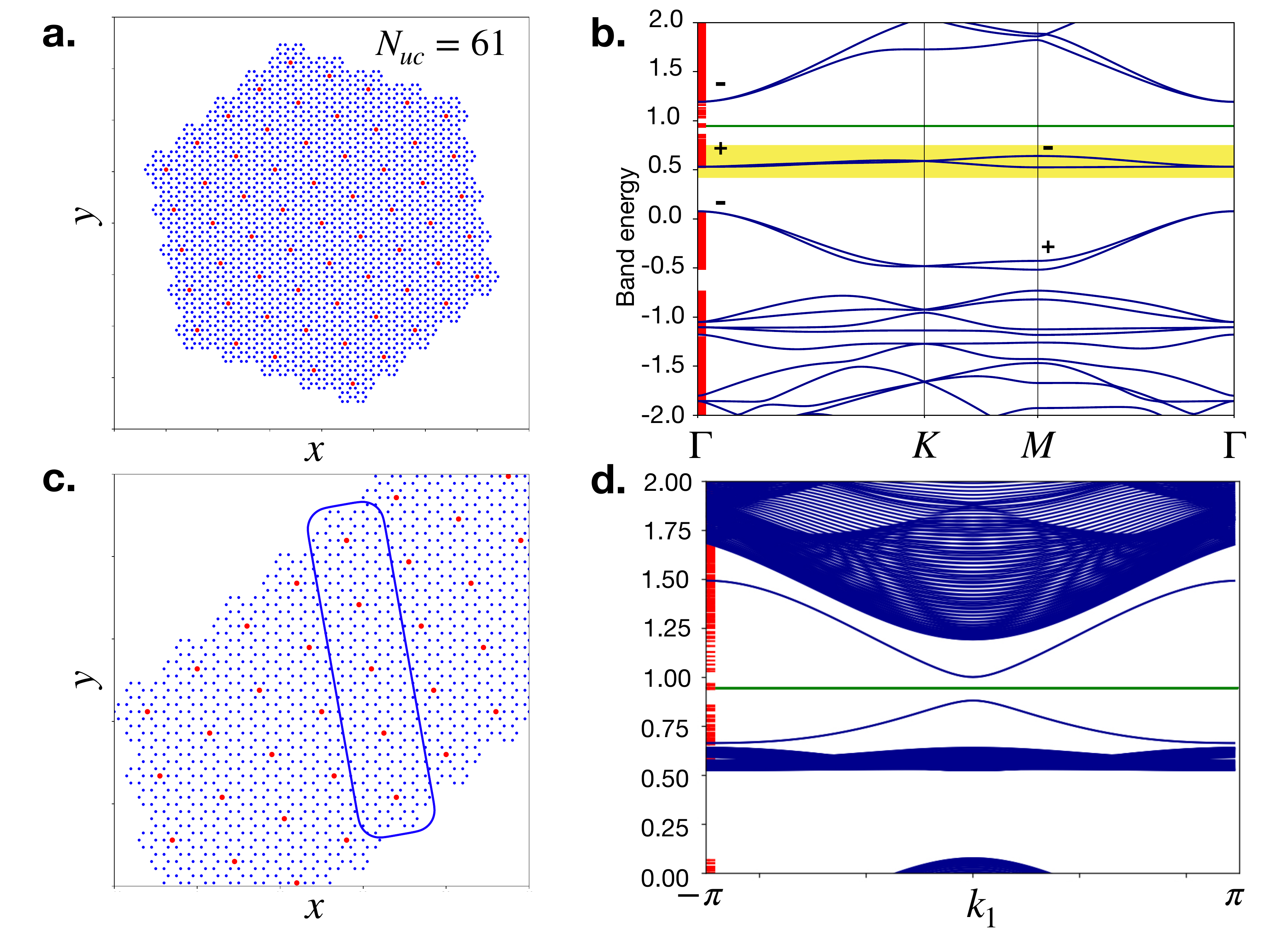}
  \caption{\label{fig:pbc_obc_tbmodel}{\bf Tight-binding calculations for open geometries} ({\bf a}) The geometry of $C_6$-symmetric flake with the number of the unit cells $N_{uc} = 61$. ({\bf b}) Band structure of the tight-binding model (in blue) with flat bands highlighted in yellow and  $C_2$-symmetry eigenvalues at $\Gamma$ and M points indicating the topological nature of the bands. The spectrum for the flake (in red) shows in-gap states around the energy indicated by the green line. ({\bf c}) The unit cell of the ribbon geometry. ({\bf d}) The spectrum for ribbon geometry (in blue) shows a small gap. In the flake geometry, there are six additional states that lie in this gap (in red).}
\end{figure}

\section*{Topological Properties}

To examine the topological properties of the flat bands constructed by the superlattice, we employ a symmetry analysis within the context of topological quantum chemistry~\cite{TQC:2017}. This analysis requires the transformation properties of the wave functions that make up the flat bands at the $\Gamma$ and the $M$ points. While the decorated graphene lattice retains the $C_6$ symmetry of graphene, all mirror symmetries are broken by the adatom arrangement, such that, including translations, our system reduces to $P_6$ space-group symmetry. For the situation shown in Fig.~\ref{fig:unit_cell}d, the flat bands highlighted with yellow transform as $\Gamma_3$ and $\Gamma_5$ at the $\Gamma$ point and $M_2$ at the $M$ point. 
In particular, the $C_2$ eigenvalues associated with these irreducible representations at the $\Gamma$ and $M$ points are $\alpha_{C_2}(\Gamma) = +1$ and $\alpha_{C_2}(M) = -1$. These eigenvalues can be understood by inspecting the orbital content of the bands in Fig.~\ref{fig:unit_cell}d: At the $\Gamma$ point, the bands stem from $p_z$ orbitals, while at the $M$ point, the bands originate from $d_{xz}$ and $d_{yz}$ orbitals.
Such a combination of irreducible representations cannot arise from an atomic limit of exponentially-localized Wannier functions~\cite{cryst1:2011,cryst2:2006,cryst3:2006}, implying that the bands are indeed topological. Following the terminology introduced in this context, these bands do not form an elementary band representation (EBR). The $p_z$-$d_{xz}$/$d_{yz}$ hybridization is thus the crucial ingredient for the formation of bands with non-trivial topology.

While the flat bands cannot be adiabatically connected to an atomic limit, they can be written as a difference between two EBRs (with integer coefficients), which indicates the fragile nature of their topology \cite{bradlyn:2019}. In particular, the bands can be expressed as the difference $\text{FT} = \text{AL}_1 - \text{AL}_2$, where $\text{AL}_1 = [\Gamma_1 \oplus \Gamma_3\Gamma_5,M_1\oplus2M_2]$ and $\text{AL}_2 = [\Gamma_1 \oplus M_1]$ are two sets of band representations forming an EBR.
This feature distinguishes the current case from a strong topological phase, which cannot be trivialised by adding trivial degrees of freedom.

Unlike strong topological phases, which are characterized by topological edge states due to the bulk-boundary correspondence, bands with fragile topology protected by $C_n$-symmetry can exhibit a filling anomaly. This topological feature describes the situation, when a mismatch exists between the number of electrons required to simultaneously satisfy charge neutrality, a unique ground state in open boundary conditions, and the crystalline symmetry. In the spectrum, $n$ degenerate states associated with the filling anomaly appear in the gap with $n/2$ states occupied for charge neutrality. In our case, adding one more electron will lead to a quantized excess charge of $e/6$ in each corner of the flake \cite{huges:2019}. 

\begin{figure}[b]
  \centering
  \includegraphics[width=1.\linewidth]{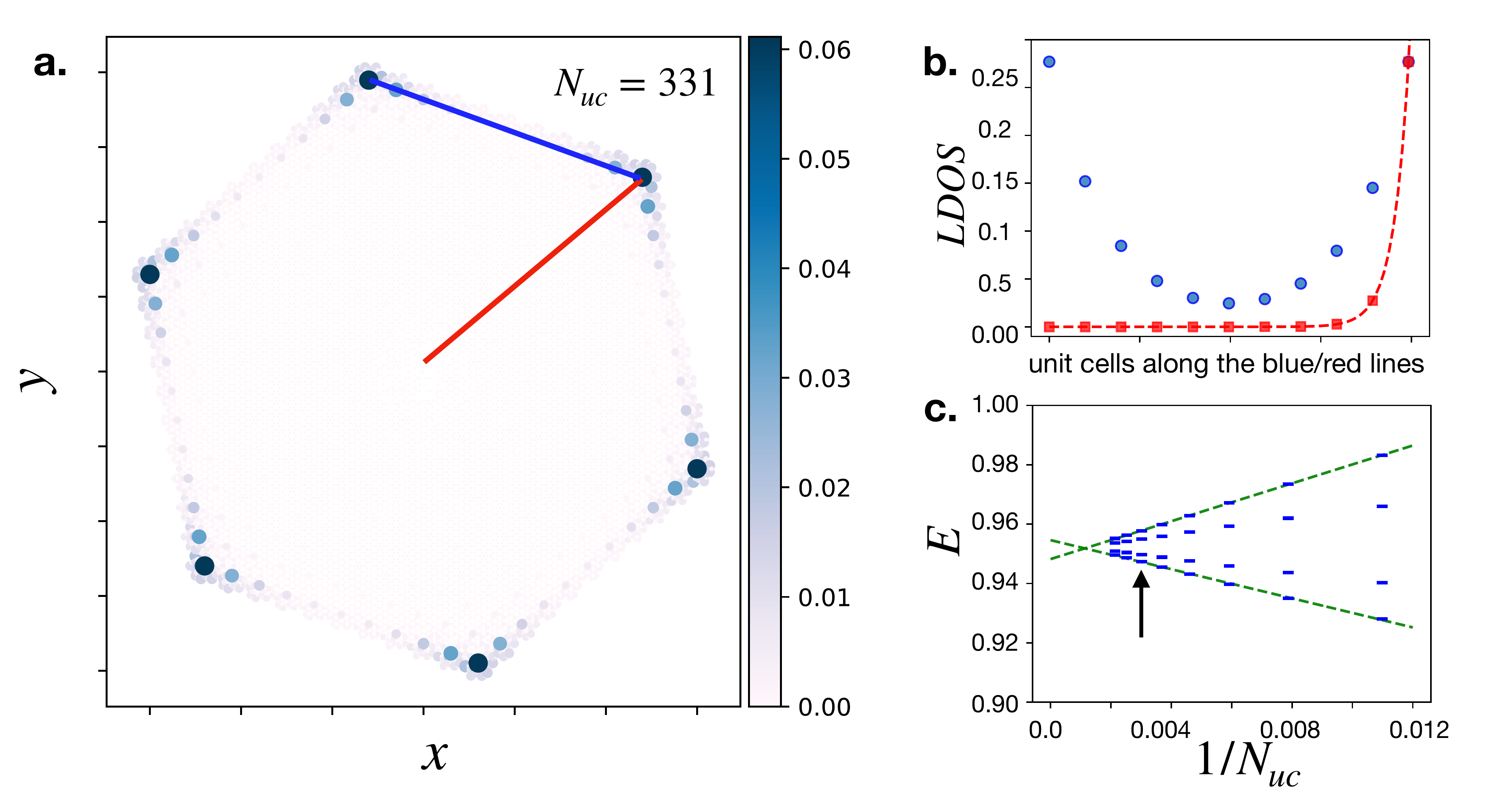}
  \caption{\label{fig:corner_states} {\bf Corner-localized in-gap states} ({\bf a}) The spatial distribution of the absolute values squared of the eigenvectors corresponding to the in-gap states for $C_6$-symmetric graphene flake shown in Fig.~\ref{fig:pbc_obc_tbmodel}a with $N_{\text{uc}}=331$. ({\bf b}) Line profile of the local density of states at the bound state, integrated along the blue and red lines given in {\bf a}. Dashed red line indicates exponential fit.  ({\bf c}) Finite-size scaling of the in-gap states showing their degeneracy in the thermodynamic limit. An arrow indicates the states with spatial distribution given in {\bf a}.}
\end{figure}

To investigate the appearance of this type of bulk-boundary correspondence associated with the fragile topology of the flat bands, we introduce a tight-binding model starting from the C $p_z$ and the transition-metal $d$ orbitals (see Methods). Such a model allows for the simulation of any open geometry, such as the ones shown in Fig.~\ref{fig:pbc_obc_tbmodel}. Before investigating this finite system further, we note that for appropriate parameters, the tight-binding model indeed yields flat bands with the correct irreducible representations as discussed above and shown in Fig.~\ref{fig:pbc_obc_tbmodel}.

The example of the $C_6$-symmetric flake in Figure~\ref{fig:pbc_obc_tbmodel}a leads to the spectrum in panel b with the spectrum of the translationally invariant system added for comparison. Indeed, we find in-gap states, though, more states than the six anticipated from fragile topology. We can understand the origin of these additional states considering a ribbon geometry as shown in Fig.~\ref{fig:pbc_obc_tbmodel}c. As can be seen in Fig.~\ref{fig:pbc_obc_tbmodel}d, most of the in-gap states are associated with edge states which are, however, not completely gapless. These states are connected to an only weakly-broken mirror symmetry perpendicular to the open direction in the ribbon geometry. In the hybridization gap of these edge states, we find six in-gap states for the flake geometry, which we attribute to the fragile topology of the system.

Figure~\ref{fig:corner_states}a shows the contribution of these six in-gap states to the local density of states of the flake geometry. Their dominant spectral weight is localized at the corners of the flake, as further emphasized in Fig.~\ref{fig:corner_states}b, which shows the unit-cell-averaged weight of the wave functions of the states in the gap along the edge. Relevant to their experimental discovery is an exponentially decaying LDOS towards the center of the structure with a characteristic length of $0.292/N_{\rm uc}$, which would allow for a distinction with respect to other edge-state observations.  Finally, while the six states associated with the filling anomaly are not degenerate in a finite geometry, a finite size scaling, Fig.~\ref{fig:corner_states}c, shows that they indeed become degenerate in the thermodynamic limit. As such, panel c in Fig.~\ref{fig:corner_states} serves as a useful guide for how the degeneracy of the corner-localized gap states evolves in the gradual buildup of such a structure.

\section*{Discussion}
The system we propose here is conceptually simple, yet features intriguing topological properties. We demonstrated how the fragile topology manifests itself through a filling anomaly, which can be mapped with a local scanning probe.
In addition to the finite geometries that are required to probe the corner-localized states, our approach allows for more design freedom. In particular, while we focused here on a superstructure with $P_6$ symmetry, any subgroup of the graphene space group $P_6/mmm$ can be realized by choosing the appropriate superlattice vectors. Furthermore, defects in the lattice, which are a distinct way of probing topological bands, can be readily implemented by the deliberate addition or removal of atoms.

Our ideas of engineering topologically non-trivial flat bands through nanostructuring go beyond the periodic decoration of graphene with adatoms. A further promising route towards their realization can be based on artificial graphene, for example using scanning-tunneling-microscopic methods to arrange CO molecules on a Cu(111) surface~\cite{gomes:2012}.
Finally, in view of the required nanometer periodicity, we expect that graphene sheets could even be engineered by lithography techniques.

\section*{Acknowledgements}
A.S., S.S.T. and T.N. were supported by funding from the European Research Council (ERC) under the European Union's Horizon 2020 research and innovation program (ERC-StG-Neupert-757867-PARATOP). A.S. was also supported by Forschungskredit of the University of Zurich, grant No. FK-20-101. S.S.T. and T.N. were additionally supported by NCRR Marvel.  S.S.T. also acknowledges support from the Swiss National Science Foundation (grant number: PP00P2\_176877). F.D.N. thanks SNSF (PP00P2-176866) and ONR (N00014-20-1-2352) for generous support.

\bibliography{biblio}

\begin{thebibliography}{42}%
\makeatletter
\providecommand \@ifxundefined [1]{%
 \@ifx{#1\undefined}
}%
\providecommand \@ifnum [1]{%
 \ifnum #1\expandafter \@firstoftwo
 \else \expandafter \@secondoftwo
 \fi
}%
\providecommand \@ifx [1]{%
 \ifx #1\expandafter \@firstoftwo
 \else \expandafter \@secondoftwo
 \fi
}%
\providecommand \natexlab [1]{#1}%
\providecommand \enquote  [1]{``#1''}%
\providecommand \bibnamefont  [1]{#1}%
\providecommand \bibfnamefont [1]{#1}%
\providecommand \citenamefont [1]{#1}%
\providecommand \href@noop [0]{\@secondoftwo}%
\providecommand \href [0]{\begingroup \@sanitize@url \@href}%
\providecommand \@href[1]{\@@startlink{#1}\@@href}%
\providecommand \@@href[1]{\endgroup#1\@@endlink}%
\providecommand \@sanitize@url [0]{\catcode `\\12\catcode `\$12\catcode
  `\&12\catcode `\#12\catcode `\^12\catcode `\_12\catcode `\%12\relax}%
\providecommand \@@startlink[1]{}%
\providecommand \@@endlink[0]{}%
\providecommand \url  [0]{\begingroup\@sanitize@url \@url }%
\providecommand \@url [1]{\endgroup\@href {#1}{\urlprefix }}%
\providecommand \urlprefix  [0]{URL }%
\providecommand \Eprint [0]{\href }%
\providecommand \doibase [0]{http://dx.doi.org/}%
\providecommand \selectlanguage [0]{\@gobble}%
\providecommand \bibinfo  [0]{\@secondoftwo}%
\providecommand \bibfield  [0]{\@secondoftwo}%
\providecommand \translation [1]{[#1]}%
\providecommand \BibitemOpen [0]{}%
\providecommand \bibitemStop [0]{}%
\providecommand \bibitemNoStop [0]{.\EOS\space}%
\providecommand \EOS [0]{\spacefactor3000\relax}%
\providecommand \BibitemShut  [1]{\csname bibitem#1\endcsname}%
\let\auto@bib@innerbib\@empty
%</preamble>
\bibitem [{\citenamefont {{Yan}}\ \emph {et~al.}(2020)\citenamefont {{Yan}},
  \citenamefont {{Zhu}}, \citenamefont {{Dong}}, \citenamefont {{Ding}},\ and\
  \citenamefont {{Xiao}}}]{yan:2020}%
  \BibitemOpen
  \bibfield  {author} {\bibinfo {author} {\bibfnamefont {S.}~\bibnamefont
  {{Yan}}}, \bibinfo {author} {\bibfnamefont {X.}~\bibnamefont {{Zhu}}},
  \bibinfo {author} {\bibfnamefont {J.}~\bibnamefont {{Dong}}}, \bibinfo
  {author} {\bibfnamefont {Y.}~\bibnamefont {{Ding}}}, \ and\ \bibinfo {author}
  {\bibfnamefont {S.}~\bibnamefont {{Xiao}}},\ }\href@noop {} {\bibfield
  {journal} {\bibinfo  {journal} {Nanophotonics}\ }\textbf {\bibinfo {volume}
  {9}},\ \bibinfo {pages} {74} (\bibinfo {year} {2020})}\BibitemShut {NoStop}%
\bibitem [{\citenamefont {Allan}\ \emph {et~al.}(2017)\citenamefont {Allan},
  \citenamefont {Fischer}, \citenamefont {Ostojic},\ and\ \citenamefont
  {Andringa}}]{allan:2017}%
  \BibitemOpen
  \bibfield  {author} {\bibinfo {author} {\bibfnamefont {M.~P.}\ \bibnamefont
  {Allan}}, \bibinfo {author} {\bibfnamefont {M.~H.}\ \bibnamefont {Fischer}},
  \bibinfo {author} {\bibfnamefont {O.}~\bibnamefont {Ostojic}}, \ and\
  \bibinfo {author} {\bibfnamefont {A.}~\bibnamefont {Andringa}},\ }\href
  {\doibase 10.21468/SciPostPhys.3.2.010} {\bibfield  {journal} {\bibinfo
  {journal} {SciPost Phys.}\ }\textbf {\bibinfo {volume} {3}},\ \bibinfo
  {pages} {010} (\bibinfo {year} {2017})}\BibitemShut {NoStop}%
\bibitem [{\citenamefont {Grigorescu}\ and\ \citenamefont
  {Hagen}(2009)}]{grigorescu:2009}%
  \BibitemOpen
  \bibfield  {author} {\bibinfo {author} {\bibfnamefont {A.~E.}\ \bibnamefont
  {Grigorescu}}\ and\ \bibinfo {author} {\bibfnamefont {C.~W.}\ \bibnamefont
  {Hagen}},\ }\href@noop {} {\bibfield  {journal} {\bibinfo  {journal}
  {Nanotechnology}\ }\textbf {\bibinfo {volume} {20}},\ \bibinfo {pages}
  {292001} (\bibinfo {year} {2009})}\BibitemShut {NoStop}%
\bibitem [{\citenamefont {Genet}\ and\ \citenamefont
  {Ebbesen}(2007)}]{genet:2007}%
  \BibitemOpen
  \bibfield  {author} {\bibinfo {author} {\bibfnamefont {C.}~\bibnamefont
  {Genet}}\ and\ \bibinfo {author} {\bibfnamefont {T.~W.}\ \bibnamefont
  {Ebbesen}},\ }\href {\doibase 10.1038/nature05350} {\bibfield  {journal}
  {\bibinfo  {journal} {Nature}\ }\textbf {\bibinfo {volume} {445}},\ \bibinfo
  {pages} {39} (\bibinfo {year} {2007})}\BibitemShut {NoStop}%
\bibitem [{\citenamefont {Bistritzer}\ and\ \citenamefont
  {MacDonald}(2011)}]{mcdonald:2011}%
  \BibitemOpen
  \bibfield  {author} {\bibinfo {author} {\bibfnamefont {R.}~\bibnamefont
  {Bistritzer}}\ and\ \bibinfo {author} {\bibfnamefont {A.~H.}\ \bibnamefont
  {MacDonald}},\ }\href {\doibase 10.1073/pnas.1108174108} {\bibfield
  {journal} {\bibinfo  {journal} {Proceedings of the National Academy of
  Sciences}\ }\textbf {\bibinfo {volume} {108}},\ \bibinfo {pages} {12233}
  (\bibinfo {year} {2011})}\BibitemShut {NoStop}%
\bibitem [{\citenamefont {Cao}\ \emph {et~al.}(2018{\natexlab{a}})\citenamefont
  {Cao}, \citenamefont {Fatemi}, \citenamefont {Demir}, \citenamefont {Fang},
  \citenamefont {Tomarken}, \citenamefont {Luo}, \citenamefont
  {Sanchez-Yamagishi}, \citenamefont {Watanabe}, \citenamefont {Taniguchi},
  \citenamefont {Kaxiras}, \citenamefont {Ashoori},\ and\ \citenamefont
  {Jarillo-Herrero}}]{cao:2018a}%
  \BibitemOpen
  \bibfield  {author} {\bibinfo {author} {\bibfnamefont {Y.}~\bibnamefont
  {Cao}}, \bibinfo {author} {\bibfnamefont {V.}~\bibnamefont {Fatemi}},
  \bibinfo {author} {\bibfnamefont {A.}~\bibnamefont {Demir}}, \bibinfo
  {author} {\bibfnamefont {S.}~\bibnamefont {Fang}}, \bibinfo {author}
  {\bibfnamefont {S.~L.}\ \bibnamefont {Tomarken}}, \bibinfo {author}
  {\bibfnamefont {J.~Y.}\ \bibnamefont {Luo}}, \bibinfo {author} {\bibfnamefont
  {J.~D.}\ \bibnamefont {Sanchez-Yamagishi}}, \bibinfo {author} {\bibfnamefont
  {K.}~\bibnamefont {Watanabe}}, \bibinfo {author} {\bibfnamefont
  {T.}~\bibnamefont {Taniguchi}}, \bibinfo {author} {\bibfnamefont
  {E.}~\bibnamefont {Kaxiras}}, \bibinfo {author} {\bibfnamefont {R.~C.}\
  \bibnamefont {Ashoori}}, \ and\ \bibinfo {author} {\bibfnamefont
  {P.}~\bibnamefont {Jarillo-Herrero}},\ }\href@noop {} {\bibfield  {journal}
  {\bibinfo  {journal} {Nature}\ }\textbf {\bibinfo {volume} {556}},\ \bibinfo
  {pages} {80} (\bibinfo {year} {2018}{\natexlab{a}})}\BibitemShut {NoStop}%
\bibitem [{\citenamefont {Cao}\ \emph {et~al.}(2018{\natexlab{b}})\citenamefont
  {Cao}, \citenamefont {Fatemi}, \citenamefont {Fang}, \citenamefont
  {Watanabe}, \citenamefont {Taniguchi}, \citenamefont {Kaxiras},\ and\
  \citenamefont {Jarillo-Herrero}}]{herrero:2018}%
  \BibitemOpen
  \bibfield  {author} {\bibinfo {author} {\bibfnamefont {Y.}~\bibnamefont
  {Cao}}, \bibinfo {author} {\bibfnamefont {V.}~\bibnamefont {Fatemi}},
  \bibinfo {author} {\bibfnamefont {S.}~\bibnamefont {Fang}}, \bibinfo {author}
  {\bibfnamefont {K.}~\bibnamefont {Watanabe}}, \bibinfo {author}
  {\bibfnamefont {T.}~\bibnamefont {Taniguchi}}, \bibinfo {author}
  {\bibfnamefont {E.}~\bibnamefont {Kaxiras}}, \ and\ \bibinfo {author}
  {\bibfnamefont {P.}~\bibnamefont {Jarillo-Herrero}},\ }\href@noop {}
  {\bibfield  {journal} {\bibinfo  {journal} {Nature}\ }\textbf {\bibinfo
  {volume} {556}},\ \bibinfo {pages} {43} (\bibinfo {year}
  {2018}{\natexlab{b}})}\BibitemShut {NoStop}%
\bibitem [{\citenamefont {Zou}\ \emph {et~al.}(2018)\citenamefont {Zou},
  \citenamefont {Po}, \citenamefont {Vishwanath},\ and\ \citenamefont
  {Senthil}}]{zou:2020}%
  \BibitemOpen
  \bibfield  {author} {\bibinfo {author} {\bibfnamefont {L.}~\bibnamefont
  {Zou}}, \bibinfo {author} {\bibfnamefont {H.~C.}\ \bibnamefont {Po}},
  \bibinfo {author} {\bibfnamefont {A.}~\bibnamefont {Vishwanath}}, \ and\
  \bibinfo {author} {\bibfnamefont {T.}~\bibnamefont {Senthil}},\ }\href
  {\doibase 10.1103/PhysRevB.98.085435} {\bibfield  {journal} {\bibinfo
  {journal} {Phys. Rev. B}\ }\textbf {\bibinfo {volume} {98}},\ \bibinfo
  {pages} {085435} (\bibinfo {year} {2018})}\BibitemShut {NoStop}%
\bibitem [{\citenamefont {{Song}}\ \emph {et~al.}(2020)\citenamefont {{Song}},
  \citenamefont {{Lian}}, \citenamefont {{Regnault}},\ and\ \citenamefont
  {{Bernevig}}}]{bernevig:2020}%
  \BibitemOpen
  \bibfield  {author} {\bibinfo {author} {\bibfnamefont {Z.-D.}\ \bibnamefont
  {{Song}}}, \bibinfo {author} {\bibfnamefont {B.}~\bibnamefont {{Lian}}},
  \bibinfo {author} {\bibfnamefont {N.}~\bibnamefont {{Regnault}}}, \ and\
  \bibinfo {author} {\bibfnamefont {B.~A.}\ \bibnamefont {{Bernevig}}},\
  }\href@noop {} {\bibfield  {journal} {\bibinfo  {journal} {arXiv e-prints}\
  ,\ \bibinfo {eid} {arXiv:2009.11872}} (\bibinfo {year} {2020})}\BibitemShut
  {NoStop}%
\bibitem [{\citenamefont {Lu}\ \emph {et~al.}(2019)\citenamefont {Lu},
  \citenamefont {Stepanov}, \citenamefont {Yang}, \citenamefont {Xie},\ and\
  \citenamefont {Efetov}}]{efetov:2019}%
  \BibitemOpen
  \bibfield  {author} {\bibinfo {author} {\bibfnamefont {X.}~\bibnamefont
  {Lu}}, \bibinfo {author} {\bibfnamefont {P.}~\bibnamefont {Stepanov}},
  \bibinfo {author} {\bibfnamefont {W.}~\bibnamefont {Yang}}, \bibinfo {author}
  {\bibfnamefont {M.}~\bibnamefont {Xie}}, \ and\ \bibinfo {author}
  {\bibfnamefont {D.~K.}\ \bibnamefont {Efetov}},\ }\href@noop {} {\bibfield
  {journal} {\bibinfo  {journal} {Nature}\ ,\ \bibinfo {pages} {653}} (\bibinfo
  {year} {2019})}\BibitemShut {NoStop}%
\bibitem [{\citenamefont {{Wang}}\ \emph {et~al.}(2021)\citenamefont {{Wang}},
  \citenamefont {{Herzog-Arbeitman}}, \citenamefont {{Burg}}, \citenamefont
  {{Zhu}}, \citenamefont {{Watanabe}}, \citenamefont {{Taniguchi}},
  \citenamefont {{MacDonald}}, \citenamefont {{Bernevig}},\ and\ \citenamefont
  {{Tutuc}}}]{Wang:2021}%
  \BibitemOpen
  \bibfield  {author} {\bibinfo {author} {\bibfnamefont {Y.}~\bibnamefont
  {{Wang}}}, \bibinfo {author} {\bibfnamefont {J.}~\bibnamefont
  {{Herzog-Arbeitman}}}, \bibinfo {author} {\bibfnamefont {G.~W.}\ \bibnamefont
  {{Burg}}}, \bibinfo {author} {\bibfnamefont {J.}~\bibnamefont {{Zhu}}},
  \bibinfo {author} {\bibfnamefont {K.}~\bibnamefont {{Watanabe}}}, \bibinfo
  {author} {\bibfnamefont {T.}~\bibnamefont {{Taniguchi}}}, \bibinfo {author}
  {\bibfnamefont {A.~H.}\ \bibnamefont {{MacDonald}}}, \bibinfo {author}
  {\bibfnamefont {B.~A.}\ \bibnamefont {{Bernevig}}}, \ and\ \bibinfo {author}
  {\bibfnamefont {E.}~\bibnamefont {{Tutuc}}},\ }\href@noop {} {\bibfield
  {journal} {\bibinfo  {journal} {arXiv e-prints}\ ,\ \bibinfo {eid}
  {arXiv:2101.03621}} (\bibinfo {year} {2021})}\BibitemShut {NoStop}%
\bibitem [{\citenamefont {{Pierce}}\ \emph {et~al.}(2021)\citenamefont
  {{Pierce}}, \citenamefont {{Xie}}, \citenamefont {{Park}}, \citenamefont
  {{Khalaf}}, \citenamefont {{Lee}}, \citenamefont {{Cao}}, \citenamefont
  {{Parker}}, \citenamefont {{Forrester}}, \citenamefont {{Chen}},
  \citenamefont {{Watanabe}}, \citenamefont {{Taniguchi}}, \citenamefont
  {{Vishwanath}}, \citenamefont {{Jarillo-Herrero}},\ and\ \citenamefont
  {{Yacoby}}}]{pierce:2021}%
  \BibitemOpen
  \bibfield  {author} {\bibinfo {author} {\bibfnamefont {A.~T.}\ \bibnamefont
  {{Pierce}}}, \bibinfo {author} {\bibfnamefont {Y.}~\bibnamefont {{Xie}}},
  \bibinfo {author} {\bibfnamefont {J.~M.}\ \bibnamefont {{Park}}}, \bibinfo
  {author} {\bibfnamefont {E.}~\bibnamefont {{Khalaf}}}, \bibinfo {author}
  {\bibfnamefont {S.~H.}\ \bibnamefont {{Lee}}}, \bibinfo {author}
  {\bibfnamefont {Y.}~\bibnamefont {{Cao}}}, \bibinfo {author} {\bibfnamefont
  {D.~E.}\ \bibnamefont {{Parker}}}, \bibinfo {author} {\bibfnamefont {P.~R.}\
  \bibnamefont {{Forrester}}}, \bibinfo {author} {\bibfnamefont
  {S.}~\bibnamefont {{Chen}}}, \bibinfo {author} {\bibfnamefont
  {K.}~\bibnamefont {{Watanabe}}}, \bibinfo {author} {\bibfnamefont
  {T.}~\bibnamefont {{Taniguchi}}}, \bibinfo {author} {\bibfnamefont
  {A.}~\bibnamefont {{Vishwanath}}}, \bibinfo {author} {\bibfnamefont
  {P.}~\bibnamefont {{Jarillo-Herrero}}}, \ and\ \bibinfo {author}
  {\bibfnamefont {A.}~\bibnamefont {{Yacoby}}},\ }\href@noop {} {\bibfield
  {journal} {\bibinfo  {journal} {arXiv e-prints}\ ,\ \bibinfo {eid}
  {arXiv:2101.04123}} (\bibinfo {year} {2021})}\BibitemShut {NoStop}%
\bibitem [{\citenamefont {Serlin}\ \emph {et~al.}(2020)\citenamefont {Serlin},
  \citenamefont {Tschirhart}, \citenamefont {Polshyn}, \citenamefont {Zhang},
  \citenamefont {Zhu}, \citenamefont {Watanabe}, \citenamefont {Taniguchi},
  \citenamefont {Balents},\ and\ \citenamefont {Young}}]{serlin:2020}%
  \BibitemOpen
  \bibfield  {author} {\bibinfo {author} {\bibfnamefont {M.}~\bibnamefont
  {Serlin}}, \bibinfo {author} {\bibfnamefont {C.~L.}\ \bibnamefont
  {Tschirhart}}, \bibinfo {author} {\bibfnamefont {H.}~\bibnamefont {Polshyn}},
  \bibinfo {author} {\bibfnamefont {Y.}~\bibnamefont {Zhang}}, \bibinfo
  {author} {\bibfnamefont {J.}~\bibnamefont {Zhu}}, \bibinfo {author}
  {\bibfnamefont {K.}~\bibnamefont {Watanabe}}, \bibinfo {author}
  {\bibfnamefont {T.}~\bibnamefont {Taniguchi}}, \bibinfo {author}
  {\bibfnamefont {L.}~\bibnamefont {Balents}}, \ and\ \bibinfo {author}
  {\bibfnamefont {A.~F.}\ \bibnamefont {Young}},\ }\href@noop {} {\bibfield
  {journal} {\bibinfo  {journal} {Science}\ }\textbf {\bibinfo {volume}
  {367}},\ \bibinfo {pages} {900} (\bibinfo {year} {2020})}\BibitemShut
  {NoStop}%
\bibitem [{\citenamefont {{Cano}}\ \emph {et~al.}(2020)\citenamefont {{Cano}},
  \citenamefont {{Fang}}, \citenamefont {{Pixley}},\ and\ \citenamefont
  {{Wilson}}}]{cano:2020}%
  \BibitemOpen
  \bibfield  {author} {\bibinfo {author} {\bibfnamefont {J.}~\bibnamefont
  {{Cano}}}, \bibinfo {author} {\bibfnamefont {S.}~\bibnamefont {{Fang}}},
  \bibinfo {author} {\bibfnamefont {J.~H.}\ \bibnamefont {{Pixley}}}, \ and\
  \bibinfo {author} {\bibfnamefont {J.~H.}\ \bibnamefont {{Wilson}}},\
  }\href@noop {} {\bibfield  {journal} {\bibinfo  {journal} {arXiv e-prints}\
  ,\ \bibinfo {eid} {arXiv:2010.09726}} (\bibinfo {year} {2020})}\BibitemShut
  {NoStop}%
\bibitem [{\citenamefont {{Wang}}\ \emph {et~al.}(2020)\citenamefont {{Wang}},
  \citenamefont {{Yuan}},\ and\ \citenamefont {{Fu}}}]{wang:2020}%
  \BibitemOpen
  \bibfield  {author} {\bibinfo {author} {\bibfnamefont {T.}~\bibnamefont
  {{Wang}}}, \bibinfo {author} {\bibfnamefont {N.~F.~Q.}\ \bibnamefont
  {{Yuan}}}, \ and\ \bibinfo {author} {\bibfnamefont {L.}~\bibnamefont
  {{Fu}}},\ }\href@noop {} {\bibfield  {journal} {\bibinfo  {journal} {arXiv
  e-prints}\ ,\ \bibinfo {eid} {arXiv:2010.09753}} (\bibinfo {year}
  {2020})}\BibitemShut {NoStop}%
\bibitem [{\citenamefont {Uri}\ \emph {et~al.}(2020)\citenamefont {Uri},
  \citenamefont {Grover}, \citenamefont {Cao}, \citenamefont {Crosse},
  \citenamefont {Bagani}, \citenamefont {Rodan-Legrain}, \citenamefont
  {Myasoedov}, \citenamefont {Watanabe}, \citenamefont {Taniguchi},
  \citenamefont {Moon}, \citenamefont {Koshino}, \citenamefont
  {Jarillo-Herrero},\ and\ \citenamefont {Zeldov}}]{uri:2020}%
  \BibitemOpen
  \bibfield  {author} {\bibinfo {author} {\bibfnamefont {A.}~\bibnamefont
  {Uri}}, \bibinfo {author} {\bibfnamefont {S.}~\bibnamefont {Grover}},
  \bibinfo {author} {\bibfnamefont {Y.}~\bibnamefont {Cao}}, \bibinfo {author}
  {\bibfnamefont {J.~A.}\ \bibnamefont {Crosse}}, \bibinfo {author}
  {\bibfnamefont {K.}~\bibnamefont {Bagani}}, \bibinfo {author} {\bibfnamefont
  {D.}~\bibnamefont {Rodan-Legrain}}, \bibinfo {author} {\bibfnamefont
  {Y.}~\bibnamefont {Myasoedov}}, \bibinfo {author} {\bibfnamefont
  {K.}~\bibnamefont {Watanabe}}, \bibinfo {author} {\bibfnamefont
  {T.}~\bibnamefont {Taniguchi}}, \bibinfo {author} {\bibfnamefont
  {P.}~\bibnamefont {Moon}}, \bibinfo {author} {\bibfnamefont {M.}~\bibnamefont
  {Koshino}}, \bibinfo {author} {\bibfnamefont {P.}~\bibnamefont
  {Jarillo-Herrero}}, \ and\ \bibinfo {author} {\bibfnamefont {E.}~\bibnamefont
  {Zeldov}},\ }\href@noop {} {\bibfield  {journal} {\bibinfo  {journal}
  {Nature}\ }\textbf {\bibinfo {volume} {581}},\ \bibinfo {pages} {47}
  (\bibinfo {year} {2020})}\BibitemShut {NoStop}%
\bibitem [{\citenamefont {Benschop}\ \emph {et~al.}(2020)\citenamefont
  {Benschop}, \citenamefont {de~Jong}, \citenamefont {Stepanov}, \citenamefont
  {Lu}, \citenamefont {Stalman}, \citenamefont {van~der Molen}, \citenamefont
  {Efetov},\ and\ \citenamefont {Allan}}]{benschop:2020tmp}%
  \BibitemOpen
  \bibfield  {author} {\bibinfo {author} {\bibfnamefont {T.}~\bibnamefont
  {Benschop}}, \bibinfo {author} {\bibfnamefont {T.~A.}\ \bibnamefont
  {de~Jong}}, \bibinfo {author} {\bibfnamefont {P.}~\bibnamefont {Stepanov}},
  \bibinfo {author} {\bibfnamefont {X.}~\bibnamefont {Lu}}, \bibinfo {author}
  {\bibfnamefont {V.}~\bibnamefont {Stalman}}, \bibinfo {author} {\bibfnamefont
  {S.~J.}\ \bibnamefont {van~der Molen}}, \bibinfo {author} {\bibfnamefont
  {D.~K.}\ \bibnamefont {Efetov}}, \ and\ \bibinfo {author} {\bibfnamefont
  {M.~P.}\ \bibnamefont {Allan}},\ }\href@noop {} {\bibfield  {journal}
  {\bibinfo  {journal} {arXiv e-prints}\ ,\ \bibinfo {eid} {arXiv:2008.13766}}
  (\bibinfo {year} {2020})}\BibitemShut {NoStop}%
\bibitem [{\citenamefont {Lee}\ \emph {et~al.}(2020)\citenamefont {Lee},
  \citenamefont {Geng}, \citenamefont {Park}, \citenamefont {Oshikawa},
  \citenamefont {Lee}, \citenamefont {Yeom},\ and\ \citenamefont
  {Cho}}]{lee:2020}%
  \BibitemOpen
  \bibfield  {author} {\bibinfo {author} {\bibfnamefont {J.~M.}\ \bibnamefont
  {Lee}}, \bibinfo {author} {\bibfnamefont {C.}~\bibnamefont {Geng}}, \bibinfo
  {author} {\bibfnamefont {J.~W.}\ \bibnamefont {Park}}, \bibinfo {author}
  {\bibfnamefont {M.}~\bibnamefont {Oshikawa}}, \bibinfo {author}
  {\bibfnamefont {S.-S.}\ \bibnamefont {Lee}}, \bibinfo {author} {\bibfnamefont
  {H.~W.}\ \bibnamefont {Yeom}}, \ and\ \bibinfo {author} {\bibfnamefont
  {G.~Y.}\ \bibnamefont {Cho}},\ }\href@noop {} {\bibfield  {journal} {\bibinfo
   {journal} {Phys. Rev. Lett.}\ }\textbf {\bibinfo {volume} {124}},\ \bibinfo
  {pages} {137002} (\bibinfo {year} {2020})}\BibitemShut {NoStop}%
\bibitem [{\citenamefont {Brar}\ \emph {et~al.}(2011)\citenamefont {Brar},
  \citenamefont {Decker}, \citenamefont {Solowan}, \citenamefont {Wang},
  \citenamefont {Maserati}, \citenamefont {Chan}, \citenamefont {Lee},
  \citenamefont {Girit}, \citenamefont {Zettl}, \citenamefont {Louie},
  \citenamefont {Cohen},\ and\ \citenamefont {Crommie}}]{brar:2011}%
  \BibitemOpen
  \bibfield  {author} {\bibinfo {author} {\bibfnamefont {V.~W.}\ \bibnamefont
  {Brar}}, \bibinfo {author} {\bibfnamefont {R.}~\bibnamefont {Decker}},
  \bibinfo {author} {\bibfnamefont {H.-M.}\ \bibnamefont {Solowan}}, \bibinfo
  {author} {\bibfnamefont {Y.}~\bibnamefont {Wang}}, \bibinfo {author}
  {\bibfnamefont {L.}~\bibnamefont {Maserati}}, \bibinfo {author}
  {\bibfnamefont {K.~T.}\ \bibnamefont {Chan}}, \bibinfo {author}
  {\bibfnamefont {H.}~\bibnamefont {Lee}}, \bibinfo {author} {\bibfnamefont
  {{\c{C}}.~O.}\ \bibnamefont {Girit}}, \bibinfo {author} {\bibfnamefont
  {A.}~\bibnamefont {Zettl}}, \bibinfo {author} {\bibfnamefont {S.~G.}\
  \bibnamefont {Louie}}, \bibinfo {author} {\bibfnamefont {M.~L.}\ \bibnamefont
  {Cohen}}, \ and\ \bibinfo {author} {\bibfnamefont {M.~F.}\ \bibnamefont
  {Crommie}},\ }\href {\doibase 10.1038/nphys1807} {\bibfield  {journal}
  {\bibinfo  {journal} {Nature Physics}\ }\textbf {\bibinfo {volume} {7}},\
  \bibinfo {pages} {43} (\bibinfo {year} {2011})}\BibitemShut {NoStop}%
\bibitem [{\citenamefont {Wang}\ \emph {et~al.}(2013)\citenamefont {Wang},
  \citenamefont {Wong}, \citenamefont {Shytov}, \citenamefont {Brar},
  \citenamefont {Choi}, \citenamefont {Wu}, \citenamefont {Tsai}, \citenamefont
  {Regan}, \citenamefont {Zettl}, \citenamefont {Kawakami}, \citenamefont
  {Louie}, \citenamefont {Levitov},\ and\ \citenamefont {Crommie}}]{wang:2013}%
  \BibitemOpen
  \bibfield  {author} {\bibinfo {author} {\bibfnamefont {Y.}~\bibnamefont
  {Wang}}, \bibinfo {author} {\bibfnamefont {D.}~\bibnamefont {Wong}}, \bibinfo
  {author} {\bibfnamefont {A.~V.}\ \bibnamefont {Shytov}}, \bibinfo {author}
  {\bibfnamefont {V.~W.}\ \bibnamefont {Brar}}, \bibinfo {author}
  {\bibfnamefont {S.}~\bibnamefont {Choi}}, \bibinfo {author} {\bibfnamefont
  {Q.}~\bibnamefont {Wu}}, \bibinfo {author} {\bibfnamefont {H.-Z.}\
  \bibnamefont {Tsai}}, \bibinfo {author} {\bibfnamefont {W.}~\bibnamefont
  {Regan}}, \bibinfo {author} {\bibfnamefont {A.}~\bibnamefont {Zettl}},
  \bibinfo {author} {\bibfnamefont {R.~K.}\ \bibnamefont {Kawakami}}, \bibinfo
  {author} {\bibfnamefont {S.~G.}\ \bibnamefont {Louie}}, \bibinfo {author}
  {\bibfnamefont {L.~S.}\ \bibnamefont {Levitov}}, \ and\ \bibinfo {author}
  {\bibfnamefont {M.~F.}\ \bibnamefont {Crommie}},\ }\href {\doibase
  10.1126/science.1234320} {\bibfield  {journal} {\bibinfo  {journal}
  {Science}\ }\textbf {\bibinfo {volume} {340}},\ \bibinfo {pages} {734}
  (\bibinfo {year} {2013})}\BibitemShut {NoStop}%
\bibitem [{\citenamefont {Wyrick}\ \emph {et~al.}(2016)\citenamefont {Wyrick},
  \citenamefont {Natterer}, \citenamefont {Zhao}, \citenamefont {Watanabe},
  \citenamefont {Taniguchi}, \citenamefont {Cullen}, \citenamefont {Zhitenev},\
  and\ \citenamefont {Stroscio}}]{wyrick:2016}%
  \BibitemOpen
  \bibfield  {author} {\bibinfo {author} {\bibfnamefont {J.}~\bibnamefont
  {Wyrick}}, \bibinfo {author} {\bibfnamefont {F.~D.}\ \bibnamefont
  {Natterer}}, \bibinfo {author} {\bibfnamefont {Y.}~\bibnamefont {Zhao}},
  \bibinfo {author} {\bibfnamefont {K.}~\bibnamefont {Watanabe}}, \bibinfo
  {author} {\bibfnamefont {T.}~\bibnamefont {Taniguchi}}, \bibinfo {author}
  {\bibfnamefont {W.~G.}\ \bibnamefont {Cullen}}, \bibinfo {author}
  {\bibfnamefont {N.~B.}\ \bibnamefont {Zhitenev}}, \ and\ \bibinfo {author}
  {\bibfnamefont {J.~A.}\ \bibnamefont {Stroscio}},\ }\href {\doibase
  10.1021/acsnano.6b05823} {\bibfield  {journal} {\bibinfo  {journal} {ACS
  Nano}\ }\textbf {\bibinfo {volume} {10}},\ \bibinfo {pages} {10698} (\bibinfo
  {year} {2016})}\BibitemShut {NoStop}%
\bibitem [{\citenamefont {Gomes}\ \emph {et~al.}(2012)\citenamefont {Gomes},
  \citenamefont {Mar}, \citenamefont {Ko}, \citenamefont {Guinea},\ and\
  \citenamefont {Manoharan}}]{gomes:2012}%
  \BibitemOpen
  \bibfield  {author} {\bibinfo {author} {\bibfnamefont {K.~K.}\ \bibnamefont
  {Gomes}}, \bibinfo {author} {\bibfnamefont {W.}~\bibnamefont {Mar}}, \bibinfo
  {author} {\bibfnamefont {W.}~\bibnamefont {Ko}}, \bibinfo {author}
  {\bibfnamefont {F.}~\bibnamefont {Guinea}}, \ and\ \bibinfo {author}
  {\bibfnamefont {H.~C.}\ \bibnamefont {Manoharan}},\ }\href@noop {} {\bibfield
   {journal} {\bibinfo  {journal} {Nature}\ }\textbf {\bibinfo {volume}
  {483}},\ \bibinfo {pages} {306} (\bibinfo {year} {2012})}\BibitemShut
  {NoStop}%
\bibitem [{\citenamefont {Robert~Drost}\ and\ \citenamefont
  {Liljeroth}(2017)}]{Drost:2017}%
  \BibitemOpen
  \bibfield  {author} {\bibinfo {author} {\bibfnamefont {A.~H.}\ \bibnamefont
  {Robert~Drost}, \bibfnamefont {Teemu~Ojanen}}\ and\ \bibinfo {author}
  {\bibfnamefont {P.}~\bibnamefont {Liljeroth}},\ }\href@noop {} {\bibfield
  {journal} {\bibinfo  {journal} {Nature Physics}\ ,\ \bibinfo {pages} {668}}
  (\bibinfo {year} {2017})}\BibitemShut {NoStop}%
\bibitem [{\citenamefont {Yan}\ and\ \citenamefont
  {Liljeroth}(2019)}]{Yan:2019}%
  \BibitemOpen
  \bibfield  {author} {\bibinfo {author} {\bibfnamefont {L.}~\bibnamefont
  {Yan}}\ and\ \bibinfo {author} {\bibfnamefont {P.}~\bibnamefont
  {Liljeroth}},\ }\href {\doibase 10.1080/23746149.2019.1651672} {\bibfield
  {journal} {\bibinfo  {journal} {Advances in Physics: X}\ }\textbf {\bibinfo
  {volume} {4}},\ \bibinfo {pages} {1651672} (\bibinfo {year}
  {2019})}\BibitemShut {NoStop}%
\bibitem [{\citenamefont {Khajetoorians}\ \emph {et~al.}(2019)\citenamefont
  {Khajetoorians}, \citenamefont {Wegner}, \citenamefont {Otte},\ and\
  \citenamefont {Swart}}]{khajetoorians:2019}%
  \BibitemOpen
  \bibfield  {author} {\bibinfo {author} {\bibfnamefont {A.~A.}\ \bibnamefont
  {Khajetoorians}}, \bibinfo {author} {\bibfnamefont {D.}~\bibnamefont
  {Wegner}}, \bibinfo {author} {\bibfnamefont {A.~F.}\ \bibnamefont {Otte}}, \
  and\ \bibinfo {author} {\bibfnamefont {I.}~\bibnamefont {Swart}},\ }\href
  {\doibase 10.1038/s42254-019-0108-5} {\bibfield  {journal} {\bibinfo
  {journal} {Nature Reviews Physics}\ }\textbf {\bibinfo {volume} {1}},\
  \bibinfo {pages} {703} (\bibinfo {year} {2019})}\BibitemShut {NoStop}%
\bibitem [{\citenamefont {Dyck}\ \emph {et~al.}(2017)\citenamefont {Dyck},
  \citenamefont {Kim}, \citenamefont {Kalinin},\ and\ \citenamefont
  {Jesse}}]{dyck:2017}%
  \BibitemOpen
  \bibfield  {author} {\bibinfo {author} {\bibfnamefont {O.}~\bibnamefont
  {Dyck}}, \bibinfo {author} {\bibfnamefont {S.}~\bibnamefont {Kim}}, \bibinfo
  {author} {\bibfnamefont {S.~V.}\ \bibnamefont {Kalinin}}, \ and\ \bibinfo
  {author} {\bibfnamefont {S.}~\bibnamefont {Jesse}},\ }\href@noop {}
  {\bibfield  {journal} {\bibinfo  {journal} {Applied Physics Letters}\
  }\textbf {\bibinfo {volume} {111}},\ \bibinfo {pages} {113104} (\bibinfo
  {year} {2017})}\BibitemShut {NoStop}%
\bibitem [{\citenamefont {Bradlyn}\ \emph {et~al.}(2017)\citenamefont
  {Bradlyn}, \citenamefont {Elcoro}, \citenamefont {Cano}, \citenamefont
  {Vergniory}, \citenamefont {Wang}, \citenamefont {Felser}, \citenamefont
  {Aroyo},\ and\ \citenamefont {Bernevig}}]{TQC:2017}%
  \BibitemOpen
  \bibfield  {author} {\bibinfo {author} {\bibfnamefont {B.}~\bibnamefont
  {Bradlyn}}, \bibinfo {author} {\bibfnamefont {L.}~\bibnamefont {Elcoro}},
  \bibinfo {author} {\bibfnamefont {J.}~\bibnamefont {Cano}}, \bibinfo {author}
  {\bibfnamefont {M.~G.}\ \bibnamefont {Vergniory}}, \bibinfo {author}
  {\bibfnamefont {Z.}~\bibnamefont {Wang}}, \bibinfo {author} {\bibfnamefont
  {C.}~\bibnamefont {Felser}}, \bibinfo {author} {\bibfnamefont {M.~I.}\
  \bibnamefont {Aroyo}}, \ and\ \bibinfo {author} {\bibfnamefont {B.~A.}\
  \bibnamefont {Bernevig}},\ }\href@noop {} {\bibfield  {journal} {\bibinfo
  {journal} {Nature}\ ,\ \bibinfo {pages} {298}} (\bibinfo {year}
  {2017})}\BibitemShut {NoStop}%
\bibitem [{\citenamefont {Po}\ \emph {et~al.}(2017)\citenamefont {Po},
  \citenamefont {Vishwanath},\ and\ \citenamefont {Watanabe}}]{watanabe:2017}%
  \BibitemOpen
  \bibfield  {author} {\bibinfo {author} {\bibfnamefont {H.~C.}\ \bibnamefont
  {Po}}, \bibinfo {author} {\bibfnamefont {A.}~\bibnamefont {Vishwanath}}, \
  and\ \bibinfo {author} {\bibfnamefont {H.}~\bibnamefont {Watanabe}},\
  }\href@noop {} {\bibfield  {journal} {\bibinfo  {journal} {Nature
  Communication}\ ,\ \bibinfo {pages} {50}} (\bibinfo {year}
  {2017})}\BibitemShut {NoStop}%
\bibitem [{\citenamefont {Po}\ \emph {et~al.}(2018)\citenamefont {Po},
  \citenamefont {Watanabe},\ and\ \citenamefont {Vishwanath}}]{po:2018}%
  \BibitemOpen
  \bibfield  {author} {\bibinfo {author} {\bibfnamefont {H.~C.}\ \bibnamefont
  {Po}}, \bibinfo {author} {\bibfnamefont {H.}~\bibnamefont {Watanabe}}, \ and\
  \bibinfo {author} {\bibfnamefont {A.}~\bibnamefont {Vishwanath}},\ }\href
  {\doibase 10.1103/PhysRevLett.121.126402} {\bibfield  {journal} {\bibinfo
  {journal} {Phys. Rev. Lett.}\ }\textbf {\bibinfo {volume} {121}},\ \bibinfo
  {pages} {126402} (\bibinfo {year} {2018})}\BibitemShut {NoStop}%
\bibitem [{\citenamefont {Benalcazar}\ \emph {et~al.}(2019)\citenamefont
  {Benalcazar}, \citenamefont {Li},\ and\ \citenamefont {Hughes}}]{huges:2019}%
  \BibitemOpen
  \bibfield  {author} {\bibinfo {author} {\bibfnamefont {W.~A.}\ \bibnamefont
  {Benalcazar}}, \bibinfo {author} {\bibfnamefont {T.}~\bibnamefont {Li}}, \
  and\ \bibinfo {author} {\bibfnamefont {T.~L.}\ \bibnamefont {Hughes}},\
  }\href@noop {} {\bibfield  {journal} {\bibinfo  {journal} {Phys. Rev. B}\
  }\textbf {\bibinfo {volume} {99}},\ \bibinfo {pages} {245151} (\bibinfo
  {year} {2019})}\BibitemShut {NoStop}%
\bibitem [{\citenamefont {Nakada}\ and\ \citenamefont
  {Ishii}(2011)}]{nakada:2011}%
  \BibitemOpen
  \bibfield  {author} {\bibinfo {author} {\bibfnamefont {K.}~\bibnamefont
  {Nakada}}\ and\ \bibinfo {author} {\bibfnamefont {A.}~\bibnamefont {Ishii}},\
  }\href
  {https://www.intechopen.com/books/graphene-simulation/dft-calculation-for-adatom-adsorption-on-graphene}
  {\emph {\bibinfo {title} {DFT Calculation for Adatom Adsorption on
  Graphene}}}\ (\bibinfo  {publisher} {IntechOpen},\ \bibinfo {year}
  {2011})\BibitemShut {NoStop}%
\bibitem [{\citenamefont {Novoselov}\ \emph {et~al.}(2004)\citenamefont
  {Novoselov}, \citenamefont {Geim}, \citenamefont {Morozov}, \citenamefont
  {Jiang}, \citenamefont {Zhang}, \citenamefont {Dubonos}, \citenamefont
  {Grigorieva},\ and\ \citenamefont {Firsov}}]{Novoselov:2004}%
  \BibitemOpen
  \bibfield  {author} {\bibinfo {author} {\bibfnamefont {K.~S.}\ \bibnamefont
  {Novoselov}}, \bibinfo {author} {\bibfnamefont {A.~K.}\ \bibnamefont {Geim}},
  \bibinfo {author} {\bibfnamefont {S.~V.}\ \bibnamefont {Morozov}}, \bibinfo
  {author} {\bibfnamefont {D.}~\bibnamefont {Jiang}}, \bibinfo {author}
  {\bibfnamefont {Y.}~\bibnamefont {Zhang}}, \bibinfo {author} {\bibfnamefont
  {S.~V.}\ \bibnamefont {Dubonos}}, \bibinfo {author} {\bibfnamefont {I.~V.}\
  \bibnamefont {Grigorieva}}, \ and\ \bibinfo {author} {\bibfnamefont {A.~A.}\
  \bibnamefont {Firsov}},\ }\href {\doibase 10.1126/science.1102896} {\bibfield
   {journal} {\bibinfo  {journal} {Science}\ }\textbf {\bibinfo {volume}
  {306}},\ \bibinfo {pages} {666} (\bibinfo {year} {2004})}\BibitemShut
  {NoStop}%
\bibitem [{\citenamefont {Aroyo}\ \emph {et~al.}(2011)\citenamefont {Aroyo},
  \citenamefont {Perez-Mato}, \citenamefont {Orobengoa}, \citenamefont {Tasci},
  \citenamefont {De~La~Flor},\ and\ \citenamefont {Kirov}}]{cryst1:2011}%
  \BibitemOpen
  \bibfield  {author} {\bibinfo {author} {\bibfnamefont {M.}~\bibnamefont
  {Aroyo}}, \bibinfo {author} {\bibfnamefont {J.}~\bibnamefont {Perez-Mato}},
  \bibinfo {author} {\bibfnamefont {D.}~\bibnamefont {Orobengoa}}, \bibinfo
  {author} {\bibfnamefont {E.}~\bibnamefont {Tasci}}, \bibinfo {author}
  {\bibfnamefont {G.}~\bibnamefont {De~La~Flor}}, \ and\ \bibinfo {author}
  {\bibfnamefont {A.}~\bibnamefont {Kirov}},\ }\href
  {https://www.scopus.com/inward/record.uri?eid=2-s2.0-80955140447&partnerID=40&md5=488772b9e21d2636a3952f66ae80ae84}
  {\bibfield  {journal} {\bibinfo  {journal} {Bulgarian Chemical
  Communications}\ }\textbf {\bibinfo {volume} {43}},\ \bibinfo {pages} {183}
  (\bibinfo {year} {2011})}\BibitemShut {NoStop}%
\bibitem [{\citenamefont {Aroyo}\ \emph
  {et~al.}(2006{\natexlab{a}})\citenamefont {Aroyo}, \citenamefont {Kirov},
  \citenamefont {Capillas}, \citenamefont {Perez-Mato},\ and\ \citenamefont
  {Wondratschek}}]{cryst2:2006}%
  \BibitemOpen
  \bibfield  {author} {\bibinfo {author} {\bibfnamefont {M.~I.}\ \bibnamefont
  {Aroyo}}, \bibinfo {author} {\bibfnamefont {A.}~\bibnamefont {Kirov}},
  \bibinfo {author} {\bibfnamefont {C.}~\bibnamefont {Capillas}}, \bibinfo
  {author} {\bibfnamefont {J.~M.}\ \bibnamefont {Perez-Mato}}, \ and\ \bibinfo
  {author} {\bibfnamefont {H.}~\bibnamefont {Wondratschek}},\ }\href {\doibase
  10.1107/S0108767305040286} {\bibfield  {journal} {\bibinfo  {journal} {Acta
  Crystallographica Section A}\ }\textbf {\bibinfo {volume} {62}},\ \bibinfo
  {pages} {115} (\bibinfo {year} {2006}{\natexlab{a}})}\BibitemShut {NoStop}%
\bibitem [{\citenamefont {Aroyo}\ \emph
  {et~al.}(2006{\natexlab{b}})\citenamefont {Aroyo}, \citenamefont
  {Perez-Mato}, \citenamefont {Capillas}, \citenamefont {Kroumova},
  \citenamefont {Ivantchev}, \citenamefont {Madariaga}, \citenamefont {Kirov},\
  and\ \citenamefont {Wondratschek}}]{cryst3:2006}%
  \BibitemOpen
  \bibfield  {author} {\bibinfo {author} {\bibfnamefont {M.~I.}\ \bibnamefont
  {Aroyo}}, \bibinfo {author} {\bibfnamefont {J.~M.}\ \bibnamefont
  {Perez-Mato}}, \bibinfo {author} {\bibfnamefont {C.}~\bibnamefont
  {Capillas}}, \bibinfo {author} {\bibfnamefont {E.}~\bibnamefont {Kroumova}},
  \bibinfo {author} {\bibfnamefont {S.}~\bibnamefont {Ivantchev}}, \bibinfo
  {author} {\bibfnamefont {G.}~\bibnamefont {Madariaga}}, \bibinfo {author}
  {\bibfnamefont {A.}~\bibnamefont {Kirov}}, \ and\ \bibinfo {author}
  {\bibfnamefont {H.}~\bibnamefont {Wondratschek}},\ }\href {\doibase
  https://doi.org/10.1524/zkri.2006.221.1.15} {\bibfield  {journal} {\bibinfo
  {journal} {Zeitschrift f{\"u}r Kristallographie - Crystalline Materials}\
  }\textbf {\bibinfo {volume} {221}},\ \bibinfo {pages} {15 } (\bibinfo {year}
  {2006}{\natexlab{b}})}\BibitemShut {NoStop}%
\bibitem [{\citenamefont {Bradlyn}\ \emph {et~al.}(2019)\citenamefont
  {Bradlyn}, \citenamefont {Wang}, \citenamefont {Cano},\ and\ \citenamefont
  {Bernevig}}]{bradlyn:2019}%
  \BibitemOpen
  \bibfield  {author} {\bibinfo {author} {\bibfnamefont {B.}~\bibnamefont
  {Bradlyn}}, \bibinfo {author} {\bibfnamefont {Z.}~\bibnamefont {Wang}},
  \bibinfo {author} {\bibfnamefont {J.}~\bibnamefont {Cano}}, \ and\ \bibinfo
  {author} {\bibfnamefont {B.~A.}\ \bibnamefont {Bernevig}},\ }\href {\doibase
  10.1103/PhysRevB.99.045140} {\bibfield  {journal} {\bibinfo  {journal} {Phys.
  Rev. B}\ }\textbf {\bibinfo {volume} {99}},\ \bibinfo {pages} {045140}
  (\bibinfo {year} {2019})}\BibitemShut {NoStop}%
\bibitem [{\citenamefont {Kresse}\ and\ \citenamefont
  {Furthm\"uller}(1996)}]{vasp}%
  \BibitemOpen
  \bibfield  {author} {\bibinfo {author} {\bibfnamefont {G.}~\bibnamefont
  {Kresse}}\ and\ \bibinfo {author} {\bibfnamefont {J.}~\bibnamefont
  {Furthm\"uller}},\ }\href {\doibase 10.1103/PhysRevB.54.11169} {\bibfield
  {journal} {\bibinfo  {journal} {Phys. Rev. B}\ }\textbf {\bibinfo {volume}
  {54}},\ \bibinfo {pages} {11169} (\bibinfo {year} {1996})}\BibitemShut
  {NoStop}%
\bibitem [{\citenamefont {Bl\"ochl}(1994)}]{PAW}%
  \BibitemOpen
  \bibfield  {author} {\bibinfo {author} {\bibfnamefont {P.~E.}\ \bibnamefont
  {Bl\"ochl}},\ }\href {\doibase 10.1103/PhysRevB.50.17953} {\bibfield
  {journal} {\bibinfo  {journal} {Phys. Rev. B}\ }\textbf {\bibinfo {volume}
  {50}},\ \bibinfo {pages} {17953} (\bibinfo {year} {1994})}\BibitemShut
  {NoStop}%
\bibitem [{\citenamefont {Kresse}\ and\ \citenamefont
  {Joubert}(1999)}]{VASPpaw}%
  \BibitemOpen
  \bibfield  {author} {\bibinfo {author} {\bibfnamefont {G.}~\bibnamefont
  {Kresse}}\ and\ \bibinfo {author} {\bibfnamefont {D.}~\bibnamefont
  {Joubert}},\ }\href {\doibase 10.1103/PhysRevB.59.1758} {\bibfield  {journal}
  {\bibinfo  {journal} {Phys. Rev. B}\ }\textbf {\bibinfo {volume} {59}},\
  \bibinfo {pages} {1758} (\bibinfo {year} {1999})}\BibitemShut {NoStop}%
\bibitem [{\citenamefont {Perdew}\ \emph {et~al.}(1996)\citenamefont {Perdew},
  \citenamefont {Burke},\ and\ \citenamefont {Ernzerhof}}]{gga-pbe}%
  \BibitemOpen
  \bibfield  {author} {\bibinfo {author} {\bibfnamefont {J.~P.}\ \bibnamefont
  {Perdew}}, \bibinfo {author} {\bibfnamefont {K.}~\bibnamefont {Burke}}, \
  and\ \bibinfo {author} {\bibfnamefont {M.}~\bibnamefont {Ernzerhof}},\ }\href
  {\doibase 10.1103/PhysRevLett.77.3865} {\bibfield  {journal} {\bibinfo
  {journal} {Phys. Rev. Lett.}\ }\textbf {\bibinfo {volume} {77}},\ \bibinfo
  {pages} {3865} (\bibinfo {year} {1996})}\BibitemShut {NoStop}%
\bibitem [{\citenamefont {Iraola}\ \emph {et~al.}(2020)\citenamefont {Iraola},
  \citenamefont {Mañes}, \citenamefont {Bradlyn}, \citenamefont {Neupert},
  \citenamefont {Vergniory},\ and\ \citenamefont {Tsirkin}}]{iraola2020irrep}%
  \BibitemOpen
  \bibfield  {author} {\bibinfo {author} {\bibfnamefont {M.}~\bibnamefont
  {Iraola}}, \bibinfo {author} {\bibfnamefont {J.~L.}\ \bibnamefont {Mañes}},
  \bibinfo {author} {\bibfnamefont {B.}~\bibnamefont {Bradlyn}}, \bibinfo
  {author} {\bibfnamefont {T.}~\bibnamefont {Neupert}}, \bibinfo {author}
  {\bibfnamefont {M.~G.}\ \bibnamefont {Vergniory}}, \ and\ \bibinfo {author}
  {\bibfnamefont {S.~S.}\ \bibnamefont {Tsirkin}},\ }\href@noop {} {\bibfield
  {journal} {\bibinfo  {journal} {arXiv e-prints}\ ,\ \bibinfo {eid}
  {arXiv:2009.01764}} (\bibinfo {year} {2020})}\BibitemShut {NoStop}%
\bibitem [{\citenamefont {Elcoro}\ \emph {et~al.}(2017)\citenamefont {Elcoro},
  \citenamefont {Bradlyn}, \citenamefont {Wang}, \citenamefont {Vergniory},
  \citenamefont {Cano}, \citenamefont {Felser}, \citenamefont {Bernevig},
  \citenamefont {Orobengoa}, \citenamefont {de~la Flor},\ and\ \citenamefont
  {Aroyo}}]{BCSDV}%
  \BibitemOpen
  \bibfield  {author} {\bibinfo {author} {\bibfnamefont {L.}~\bibnamefont
  {Elcoro}}, \bibinfo {author} {\bibfnamefont {B.}~\bibnamefont {Bradlyn}},
  \bibinfo {author} {\bibfnamefont {Z.}~\bibnamefont {Wang}}, \bibinfo {author}
  {\bibfnamefont {M.~G.}\ \bibnamefont {Vergniory}}, \bibinfo {author}
  {\bibfnamefont {J.}~\bibnamefont {Cano}}, \bibinfo {author} {\bibfnamefont
  {C.}~\bibnamefont {Felser}}, \bibinfo {author} {\bibfnamefont {B.~A.}\
  \bibnamefont {Bernevig}}, \bibinfo {author} {\bibfnamefont {D.}~\bibnamefont
  {Orobengoa}}, \bibinfo {author} {\bibfnamefont {G.}~\bibnamefont {de~la
  Flor}}, \ and\ \bibinfo {author} {\bibfnamefont {M.~I.}\ \bibnamefont
  {Aroyo}},\ }\href {\doibase 10.1107/S1600576717011712} {\bibfield  {journal}
  {\bibinfo  {journal} {Journal of Applied Crystallography}\ }\textbf {\bibinfo
  {volume} {50}},\ \bibinfo {pages} {1457} (\bibinfo {year}
  {2017})}\BibitemShut {NoStop}%
\end{thebibliography}%


\clearpage

\appendix
\section*{Methods}

\subsection{Tight-binding model}

To study low-energy properties of the described system, we build a minimal tight-binding model. For $(n,m)=(-1,2)$, we have to include 42 $p_z$ orbitals and 4 $d$ orbitals of the transition-metal adatom in the unit cell, which in a nearest-neighbour approximation give rise to the following tight-binding model:
\begin{equation}
\label{eq:tb-hamiltonian}
H = -t \sum_{\langle i,j\rangle}  \left( c^\dagger_i c_j + \text{h.c.} \right) 
+ \sum_{\alpha,\langle i,j\rangle}\left( \tilde{t}_{\alpha,j}d^\dagger_{\alpha i} c_{j}+ \text{h.c.}\right)\,,
\end{equation}
where $c^{(\dagger)}_j,d^{(\dagger)}_{\alpha j}$ correspond to creation and annihilation of $p_z$  and $d$ orbitals respectively with $\alpha \in \{xy,x^2-y^2,xz,yz\}$ and the hopping parameters $\tilde{t}_{\alpha,j}$, which depend on the type of $d$ orbital and can be simplified by taking into account local symmetries of the lattice.
\begin{figure}[h]
  \centering
  \includegraphics[width=1.\linewidth]{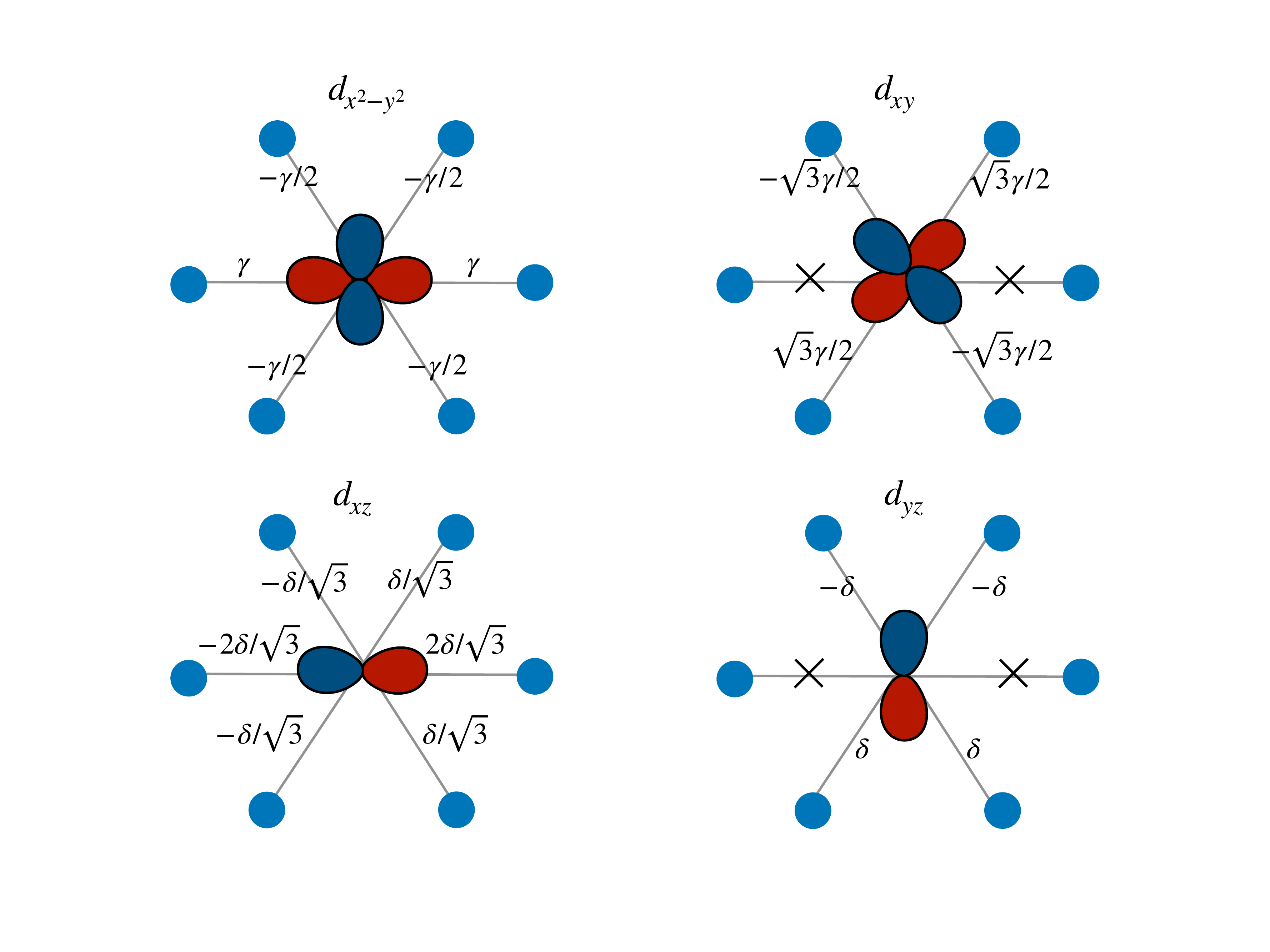}
  \caption{\label{fig:tb_hoppings} The values of the hopping parameters between four different $d$-orbitals of adatom shown in the center of each hexagon and the nearest $p_z$-orbitals of graphene. }
\end{figure}

Locally around the position of the adatom, the unit cell is invariant under mirror symmetry $\mathcal{M}_x$ and $\mathcal{C}_6$-rotation represented by 

\begin{equation}
\mathcal{C}_6 = 
\begin{pmatrix}
\begin{matrix}
0&1&0&0&0&0\\
0&0&1&0&0&0\\
0&0&0&1&0&0\\
0&0&0&0&1&0\\
0&0&0&0&0&1\\
1&0&0&0&0&0\\
\end{matrix}&\text{\huge0}\\
\text{\huge0}&\begin{matrix}R^{xz}_{yz}&\\
&R^{xy}_{x^2-y^2}
\end{matrix}
\end{pmatrix},
\end{equation}
where
\begin{equation*}
R^{xz}_{yz} = 
\begin{pmatrix}
\cos\frac{2\pi}{6}&\sin\frac{2\pi}{6}\\
-\sin\frac{2\pi}{6}&\cos\frac{2\pi}{6}\\
\end{pmatrix},
\end{equation*}

\begin{equation*}
R^{xy}_{x^2-y^2} =
\begin{pmatrix}
\cos^2(\frac{2\pi}{6}) -\sin^2(\frac{2\pi}{6})&2\cos\frac{2\pi}{6}\sin\frac{2\pi}{6}\\
2\cos\frac{2\pi}{6}\sin\frac{2\pi}{6}&\cos^2(\frac{2\pi}{6}) -\sin^2(\frac{2\pi}{6})\\
\end{pmatrix}.
\end{equation*}
and 
\begin{equation}
    \mathcal{M}_x = 
    \begin{pmatrix}
    \begin{matrix}
    0&0&0&0&0&1\\
    0&0&0&0&1&0\\
    0&0&0&1&0&0\\
    0&0&1&0&0&0\\
    0&1&0&0&0&0\\
    1&0&0&0&0&0\\
    \end{matrix}&\text{\huge0}\\
    \text{\huge0}&
    \begin{matrix}
    1&&&\\
    &-1&&\\
    &&1&\\
    &&&-1\\
    \end{matrix}
    \end{pmatrix}.
\end{equation}

The part of the Hamiltonian corresponding to the hoppings within the central hexagon including four $d$ orbitals in the center and six nearest $p_z$ orbitals surrounding them is given by the matrix
\begin{equation}
    H_{\tilde t} = \begin{cases}
t  & \text{if $i,j = \{\overline{1,6}$\},} \\
\tilde{t}_{\alpha,j} & \text{if $i = \{\overline{1,6}$\}, $j = \{\overline{7,10}$\} and vice versa,} \\
0 & \text{if $i,j = \{\overline{7,10}$\},} 
\end{cases}
\end{equation}
with unknown hopping parameters $\tilde{t}_{\alpha,j}$, which can be expressed in terms of only two parameters $\gamma$ and $\delta$ using the following set of equations provided by the symmetry constraints
\begin{equation}
\begin{split}
    \mathcal{C}_6 H_{\tilde t} {\mathcal{C}_6}^{-1} &=  H_{\tilde t}\\
    \mathcal{M}_x H_{\tilde t} {\mathcal{M}_x}^{-1} &=  H_{\tilde t}.
\end{split}
\end{equation}
The results are summarised in Fig.~\ref{fig:tb_hoppings}.

\subsection{First-principles calculations}
We performed first-principles calculations within the density functional theory (DFT) framework using the VASP package \cite{vasp}, employing the projector-augmented wave method \cite{PAW,VASPpaw} and the Perdew-Burke-Ernzerhof generalized-gradient approximation (GGA-PBE)\cite{gga-pbe} for the exchange-correlation energy. We employ the following scheme consisting of two steps: In a first step, we perform self-consistent DFT calculations on a $6\times6\times1$ grid in momentum space with fixed in-plane atomic positions and allow for the lattice relaxation in the out-of plane direction. In this way, we find the distance between the graphene lattice and adatom which minimises the total energy and we get the corresponding charge-density profile. In the second step, we use this charge density to perform non-self-consistent calculations along the $\Gamma-K-M-\Gamma$ path in momentum space to obtain the band structure plots. Note that the calculations do not include spin-orbit coupling. The irreducible representations were determined using the \texttt{IrRep} code \cite{iraola2020irrep} and given in the notation of the Bilbao Crystallographic Server \cite{BCSDV}.

\begin{figure*}[h]
  \centering
  \includegraphics[width=1.\linewidth]{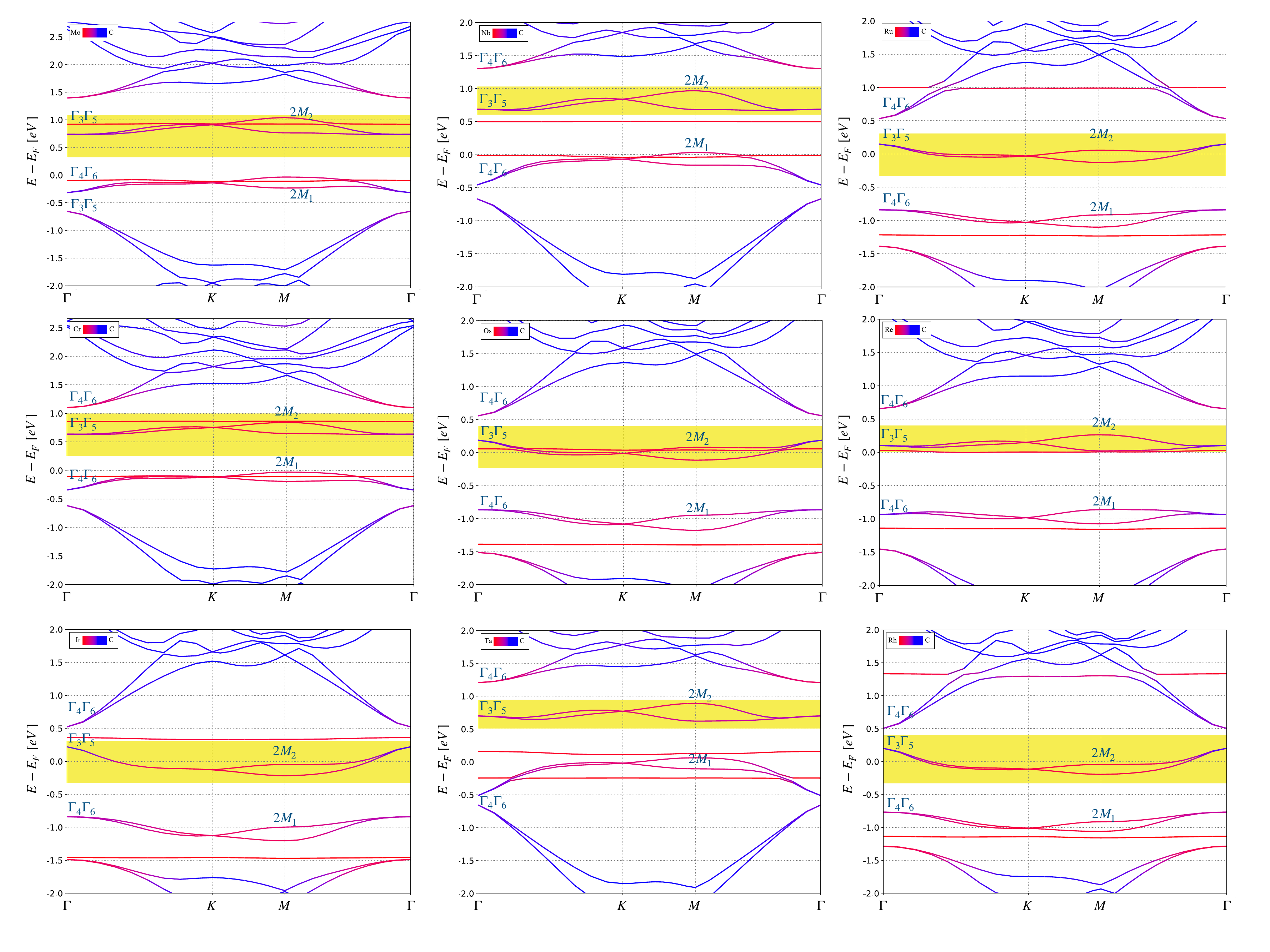}
  \caption{\label{fig:all_dft_bands} The DFT band structure with indicated irreducible representations at $\Gamma$ and $\text{M}$ momenta and highlighted flat bands featuring fragile topology for the following adatoms: Mo, Nb, Ru, Cr, Os, Re, Ir, Ta and Rh. }
\end{figure*}

Performing these calculations for different chemical elements we find a set of elements, which give rise to flat bands with fragile topology (see Fig.\ref{fig:all_dft_bands}). In Tab. \ref{tab:table-elements} we summarise the results of DFT calculations for a set of adatoms, which show similar behaviour of the band structure to what has been presented in the main text.  

\end{document}